# Global analysis of VHHs framework regions with a structural alphabet.

Floriane Noel[1,2,3,4,#,+], Alain Malpertuy[5] & Alexandre G. de Brevern[1,2,3,4,*]

[1] INSERM, U 1134, DSIMB, F-75739 Paris, France.
[2] Univ Paris Diderot, Sorbonne Paris Cité, UMR_S 1134, F-75739 Paris, France.
[3] Institut National de la Transfusion Sanguine (INTS), F-75739 Paris, France.
[4] Laboratoire d'Excellence GR-Ex, F-75739 Paris, France.
[5] Atragene, F-94200 Ivry-sur-Seine, France.
[#] present adress : Institut Curie, PSL Research University, INSERM, UMR 932, F-75005, Paris, France.
[+] present adress : Université Paris Sud, Université Paris-Saclay, F-91405 Orsay, France.

*Short title:* VHH FRs structures

* Corresponding author:
Mailing address: Dr. de Alexandre G. de Brevern, INSERM UMR_S 1134, DSIMB, Université Paris Diderot, Institut National de Transfusion Sanguine (INTS), 6, rue Alexandre Cabanel, 75739 Paris cedex 15, France
e-mail : alexandre.debrevern@univ-paris-diderot.fr

# Abstract

The VHHs are antigen-binding region/domain of camelid heavy chain antibodies (HCAb). They have many interesting biotechnological and biomedical properties due to their small size, high solubility and stability, and high affinity and specificity for their antigens. HCAb and classical IgGs are evolutionary related and share a common fold. VHHs are composed of regions considered as constant, called the frameworks (FRs) connected by Complementarity Determining Regions (CDRs), a highly variable region that provide interaction with the epitope. Actually, no systematic structural analyses had been performed on VHH structures despite a significant number of structures. This work is the first study to analyse the structural diversity of FRs of VHHs. Using a structural alphabet that allows approximating the local conformation, we show that each of the four FRs do not have a unique structure but exhibit many structural variant patterns. Moreover, no direct simple link between the local conformational change and amino acid composition can be detected. These results indicate that long-range interactions affect the local conformation of FRs and impact the building of structural models.

Keywords: secondary structure / sequence structure relationship / structural alphabet / antibodies / frameworks.

## Introduction

The antibodies (Ab) or immunoglobulins (Ig) are glycoproteins that play a central role in the immune response. They allow the recognition of antigens, the recruitment of cells and stimulation of immune defence mechanisms. They have a similar structure in all vertebrates [1]. These large molecules (~ 150 kDa) are composed of two identical heavy (H) and two identical light (L) chains, linked by disulphide bridges. The type of heavy chain of the antibody determines the type of Ig. The most common Ig is the Ig type G (IgG). These chains are arranged in variable and constant domains. The L chains are composed of a variable domain (VL) and a constant domain (CL). The H chains are composed of a variable domain (VH) and three constant domains (CH1 to CH3). Antibodies have been widely used in biotechnological applications [2]. The number of medical treatments based on the use of such macromolecules (*e.g.* oncology, infectious diseases or against autoimmune diseases) increases greatly [3-5]. However the difficulty of producing them and their costs limit their use.

In camelid family (*camelidae*), which includes camels and llamas, conventional antibodies are found, but in addition these species have particular antibodies where L chain and CH1 domain are missing. These antibodies are called Heavy Chain Antibodies (HCAbs [6]). C-terminal VH region derived from HCAbs are called VHH and Nanobody[TM]. Interestingly, even without their VL counterparts, the VHHs have an affinity and specificity at least as efficient as IgGs. VHH have also a good stability and solubility that lead to their use for biotechnological and biomedical applications [7].

These VHHs are composed of 4 regions whose sequences and structures are defined as conserved (called *Framework Regions*, *FRs*). In addition, VHHs contain three connecting regions showing high variability both in sequence content and structure conformation. These regions are complementary to the antigen surface and are called *Complementarity Determining Regions* or CDRs). Figure 1 underlines in 3D (see Figure 1A [8-16]) and 2D (see

Figure 1B) similarities of VHHs to conventional antibodies. However, some difference can be noticed in the length of the CDRs and in the FR2 residue composition. The VHHs have many interesting biotechnological properties [17]. VHHs are very small molecules (~ 15 kDa) and are very soluble and highly stable. VHHs have a high specificity and affinity for their target antigen. In addition, it is possible to humanize them by modifying few residues in FR2 [18] without altering their properties. The cloning and production of VHHs are easy to implement. In addition the VHHs are interesting alternative to the use of monoclonal antibodies for therapy as shown by recent studies of their use against the Dengue virus [19], H5N1 influenza [20], viral infection [21], aflatoxins in agro-products [22], head and neck cancers [23], vascular endothelial growth factor implicated in cancers [24], venom therapy [25] and *Plasmodium knowlesi* malaria vector [26]. Phase II clinical trials are currently underway to evaluate the efficacy and toxicity of a therapeutic Nanobody[TM] in patients with thrombotic thrombocytopenic purpura [27] showing recent promising results [28]. VHHs may also be used as support materials for the crystallography of different proteins [29], as they prevent domain mobility, can bind to interface or cavities, and also stabilize loops [21, 30]. They have been widely used for membrane proteins [31-33] and also as biosensor [34].

The structure of proteins is the support of their interactions. It is essential to have access to the VHH structures when working on their use in biotechnology. In a previous study, we highlighted the difficulty, as well as interest, to obtain a structural model of VHH directed against a specific receptor for chemokines (DARC or Duffy Antigen / Receptor for Chemokines, see [35-36]). In the present study, we focused on the VHH FR regions considered as constant. We have used all structural VHH data available from the Protein DataBank (PDB, [37-38]) to underline this *consistency*. For this purpose, we have used a structural alphabet (the Protein Blocks or BPs, [39]), which allows analysing finely the local protein structure conformations (see [40] for a review). It was used in multiple cases to analyse difference from particular proteins involved in diseases, e.g. integrins implicated into allo-immunisations [41-43] or a transmembrane receptor implicated into hypogonadism [44]. Our study shows that the FRs regions, previously considered as constant, present (unexpected) variations. We also define structural patterns for the different FRs which will help to improve the 3D structural models as the analyses of paratope / epitope.

## Materials and Methods

***Data sets***. The dataset of protein structures is taken from the Protein DataBank [37-38]. It was selected using key-words search on the PDB website [45]. VHH structures with missing residues in the structure were not taken into account. Analyses were done using an approach developed in PTM-SD [46]. The collected dataset is composed of 114 PDB files, with 160 VHH structures. Duplicated structures were deleted according to only one VHH structure was conserved if identical structures with identical sequences were identified. Then the final dataset of VHH structures is composed of 133 unique complete structures (see Sup Data 1), 70% are from *llama glama*, 4.5% from *vicugna pacos*, rest being from *camelus dromadarius*. 132 structures are X-ray structures (with a mean resolution of 2.1 Å) and one NMR structural model.

***Data analysis.*** Different analyses on VHH sequences and VHH structures were performed. The delineation of the FRs and CDRs was done using multiple alignments generated by the ClustalO software (version 1.1.0) [47]. Visualization of VHHs structures was done with the PyMOL software [48].

Scripts for the different analyses were programmed with Python language and R script language [49]. The VHH structures are superimposed with the PROFIT software [50] and mulPBA webserver [51]. Root mean square deviation (*rmsd*) was computed to compare VHH entirely or partially; *rmsd* is the square root of the average of distances (in Å) between backbone atoms of a protein structure [52].

***Secondary structure***. Assignment was performed by using the most classic approach DSSP [53] (CMBI version 4.0). DSSP assign more than four secondary structural states, thus we have reduced them as: α-helix including α, $3_{10}$ and π- helices, the β-strand containing only the β-sheet, the turn involving the turn assignments and bends, and the coil including the rest of the assignments (β-bridges and coil), as done in previous studies [54]. Default settings had been used for all methods.

The predictions of secondary structure were performed with PSIPRED [55-56] and Jpred 4 [57] software. The accuracy of the prediction is given by the $Q_3$ score which is the percentage of residues predicted in their right state (α-helix, β-strand or coil).

***Protein Blocks description.*** Protein Blocks (PBs [40]) correspond to a set of 16 local prototypes, labelled from *a* to *p* (e.g., see Figure 1 of [58]), of 5 residues length, clustered based on φ, ϕ dihedral angles description. They were obtained using an unsupervised classifier similar to Kohonen Maps [59] and Hidden Markov Models [60]. The PBs *m* and *d* can be roughly described as prototypes for central α-helix and central β-strand, respectively. PBs *a* through *c* primarily represent the N-cap region of β-strand while PBs *e* and *f* correspond to the C-caps; PBs *g* through *j* are specific to coils, *k* and *l* correspond to the N cap region of α-helix, and PBs *n* through *p* to that of C-caps. This structural alphabet allows a reasonable approximation of local protein 3D structures [39] with an average root mean square deviation (*rmsd*) of 0.42 Å [61]. PBs assignment was performed with PBxplore tool developed under the guidance of Pierre Poulain (https://github.com/pierrepo/PBxplore, *in preparation*). PBs are used to translate the 3D structures under a 1D shape like a sequence of amino acids and are particularly more accurate than the secondary structures [40, 62-63] (see Sup Data 2 to see translation of a 3D VHH into a 1D sequence).

*Analyses*. Protein Blocks were used in this study to define specific patterns associated to each FR. For the main pattern associated to one FR is simply the succession of PBs that is the most occurring. Some PBs are close so the patterns can be slightly degenerated. We define the pattern as done in Prosite, e.g. the pattern "*r* [*st*] *u*" means that at position 1, the PB seen is PB *r*, in position 2 is it a mixture of PBs *s* and *t* that are close and at position 3, the PB *u*.

Intrinsic propensity $fr_x^i$ of amino acids (or PBs) observed at a position $i$ of a sequence are normalized by dividing the frequency $f_x^i$ of this amino acid $x$ (or BP) by the observed frequency of this amino acid (or PB) in a non-redundant protein structure databank. Thus, the intrinsic propensity is $fr_x^i = \dfrac{f_x^i}{f_x^{DB}}$. If this value is one, the frequency is expected at random, if it is higher, it is over-represented and if it is less, it is under-represented.

The equivalent number of PBs [or amino acids] [39, 64] ($N_{eq}$) is a statistical measurement similar to entropy and represents the average number of PBs [or amino acids] for a given residue. $N_{eq}^i$ at position $i$ is calculated as follows: $N_{eq}^i = exp(-\sum_{x=1}^{16} f_x^i \ ln \ f_x^i)$.

A $N_{eq}$ value of 1 indicates that only one type of PB [or amino acids] is observed, while a value of 16 is equivalent to a random distribution for PBs [and 20 for amino acids].

For N-terminus FR (FR1) and C-terminus FR (FR4), it should be noted that only the positions actually found on all VHHs were analysed),

## Results

*Overall Analysis.* As noted previously in material and methods section, search in PDB

website allowed the final selection of 133 different VHH chains. As expected, both the superposition of the structures and multiple alignments showed that VHHs topology is globally well conserved with conserved regions (FRs) but and other regions less conserved (CDRs).

To summarize, the succession of FRs and CDRs is clearly detectable. However, it must be initially well defined. Then the demarcation of the FRs and CDRs of conventional antibodies (IgG) has been studied and well characterized [17] and it can be transposed to VHHs. Nonetheless, this is not a trivial process [65-68]. To define a clear separation between FRs and CDRs, a multiple alignment was done using the ClustalO software [47]. The alignment also helped to select 10 VHHs representative of the entire data set. These VHHs represent the most diverse set of VHHs i.e. with highest sequence divergence. FRs and CDRs delineation corresponds to classical IMGT numbering for VHHs [69-70].

The FRs regions are commonly considered as 'constant' regions both in terms of sequence and structure while the CDRs are variable / hypervariable [17]. The visualization of the superimposition of representative VHHs confirms that topologies of FRs are quite similar and mainly composed of β-strands (see Figure 1A). It also highlights the structural diversity of CDRs, and at lesser extend to FRs, particularly regarding the size of β-strands or the conformation of the connecting loops. We were able to simplify the topology as one 2D projection see Figure 1B), which enables a simplified analysis of VHHs. This representation is based mainly on the interaction between β-strands. Please note that it had been postulated that VHH have a CDR "4" [71]. The region between residues 71–78 (according to IMGT numbering [69-70]) is close to the other CDRs, leading to a larger paratope [72-74]. In the present study we didn't take into account this potential CDR4 as a CDR but as intrinsically part of FR3 as the sequence identity of this region was high as expected in the FR regions.

*FR analyses*. These observations led to quantify without any *a priori* this 'constancy' of FRs both in terms of sequence and of structure. Then we observed that the FRs are similar, but with some positions not conserved. Their percentages of identity range from 76.9% to 94.4% respectively for FR2 (14 residues, see Sup Data 3) and FR4 (9 residues). A specific analysis in terms of species (namely *llama glama*, *vicugna pacos*, and *camelus dromedarius*), show no specific tendencies from one species in regards to another, i.e. no sequence species specificity is observed.

We therefore analysed the 3D structure through a 1D representation via Protein Blocks. Then we clustered the succession of PBs leading to the description of FRs as patterns, this PB series can be seen as a classical Prosite patterns but not made from amino acids but with 3D information. These patterns are structural patterns as PBs represent 3D local conformation. For each FR, a distance matrix between these series of PBs has been made to group the closest structural patterns [75].

Table 1 shows the results for each FR. FR2 is the framework which has the structural pattern representing the largest number of structures (84%) while for the other FRs the most recurrent structural patterns only represents 40% (FR1), 63% (FR3) and 38% (FR4). We also observed no second highly recurrent patterns appear for each of the FRs. FR2 is the most directed structural pattern; it had the lowest number of variant patterns (5). The very long FR3 has 10 variants representatives of the remaining 37% of the structures. It is a limited number in regards to the length of this FR. No significant correlation between FR length and the percentage of structures corresponding to the variant patterns can be found, or even between this length and the number of variants patterns.

Focusing on the FRs amino acid sequence, we observed that 14 positions are constraint to one amino acid (see Figure 2). As shown in Table 1, 30 positions (out of 78) are exclusive to one type of amino acids while 14 positions show a $N_{eq}$ on amino acids of these positions quite high (> 2). Thus, for FR2 and FR3, they represent only 3 residues on 14 and 11 on 32 respectively. For FR1 and FR4, the figure is higher but associated with greater structural variability, probably due to fewer structural constraints. We also noticed that non-exclusive positions are rare and usually have very different residues.

In terms of sequence – structure relationship, we evaluate the prediction of secondary structure of the whole dataset using PSIPRED software [55-56]. PSIPRED is the most widely used secondary structure prediction method with a third version leading to an expected average prediction rate ($Q_3$) of 82%. A striking result is the weak quality of the prediction for the VHHs; the $Q_3$ value is only of 72.2%. It is slightly better for FR2 (86.4%), but lower for other the FRs with FR1 equals to 75.5%, FR3 to 66,9% and FR4 to 72.0%. This observation may be related to the difficulty to predict β-strand, *i.e.* the most difficult repetitive structure to predict [76]. Jpred 4 [57] prediction leads to a relative similar prediction rate of 78%, with similar tends. In regards to VH of IgGs, the results do not show a direct correlation.

*FR2 analyses.* The analysis of the FR2 (see Figure 3) allows us to identify a main structural pattern (*ddd*[*de*]*ehia*[*cd*]*ddfbd*) and seven variant patterns (see Table 1 and Figure 3B). The first three sub-patterns present similar structure (see Figure 3C with variant patterns FR2<sup>sm1.1</sup>, FR2<sup>sm1.2</sup> et FR2<sup>sm1.3</sup>). The central region of the main structural pattern shows a specific series of PBs *ehia*, while these variants have a variation around a motif *f*[*kb*][*bl*]*c* which are clearly distinct from *ehia* as seen in [77]. The divergence of this framework encompasses especially the loop connecting the two β-strands of FR2, showing the existence of five distinct structural subunits (having a rmsd of 1 Å which is important for this short length sequence, see 3D examples in Figure 3C).

The differences are mostly found at the same positions. The FR2<sup>sm2</sup> is associated with PB series *dfbd* which causes 'contraction' of the loop; this loop is therefore more outwardly. The FR2<sup>sm3</sup> greatly diverges as loop often associated to β-strand is replaced by a typical loop associated to helical structures, namely *fkop*. So the loop sticks up at the opposite of FR2$^{sm2}$. The PB motif *ehia* of FR2$^{sm4}$ is shifted and replace by *hiab* that is drastically different compared to the other motifs. Here the loop is much more extended, and not like any other FR2. FR2$^{sm5}$ is also different with specific features observed on the second β-strand of FR2. This allows the creation of more hydrogen bonds than other FR2 and extends the length of β-sheets.

For this framework along 14 residues, the positions 1 (residue W), 3 (residue R) and 7 (residue G) are associated to only one type of amino acids (see Figure 3A). For 4 other positions, the most represented amino acid does not exceed 80% (positions 2, 9, 12 and 14), and finally the most variable position is the position 7 with 12 residues (amino acids on $N_{eq}$> 4, Figure 3A).

Analysis of the various patterns of different FRs shows only one case of a sequence substitution clearly associated with a unique type of variant pattern, namely the FR2$^{sm3}$, characterized by the succession *fokp*, a Proline residue is observed; it is the only case observed.

*The other FRs.* Concerning FR3, which is the longest FR, the Figure 4A shows the occurrence of PBs inside the VHHs. The figure 4A underlines the variability of positions 6 to 13 in terms of structure and also in less extend the connection to CDR3. In the amino-acid sequence, 11 positions among 32 are only associated to 1 kind of residues. Five positions have equivalent residues (namely, residues are highly similar). The other positions represent half of the FR (16/32) and very different amino acid associations even with very different properties are found (see Sup Data 4).

We observed that the most variable local structure conformation is also the most variable in terms of sequence. Nonetheless, the amino acid sequences do not allow predicting the associated PBs. Only position 7 is well-conserved compare to the other positions (positions 6, 10 and 12 being associated to higher amino acid $N_{eq}$ values, see Figure 2C).

A precise analysis of the 10 variant patterns underlines that the main differences are found in positions of the sequence that are not strictly conserved in main structural pattern. In addition, no direct correlation can be seen with amino acid content and the PBs constructed sequence. As for FR3, the connecting loop is associated to various modifications (see Figure 4B). Most of them shorten the first β-strand of FR3 and we observed they are associated to high B-factors.

FR1 and FR4 are shorter than FR3 and FR2. Their main structural patterns are less represented (40 and 38% respectively). For this two FRs, only 4 positions among 23 have amino acid composition encompassing very diverse types of residues (big, aromatic, polar, charged…). This percentage is largely lower compared to FR2 and FR3. For FR1, the less determined structural region is the N-terminus region. It is also found with higher B-factors, and a low number of contacts. The more stable region starts with the second β-strand. We observed just before the second β-strand a sharp determined turn that is a PBs series *dehiac;* a very stable element that is important for protein structures [77]. The connection to CDR1, even if the length of CDR1 is short and constant, shows a high diversity in terms of local conformations. It highlights the impact of CDR1 on the final fold of FR1.

FR4 shows similar features. Its end is often short with high B-factors or unresolved, a typical case of most of the proteins [78-79]. Only the first nine positions are always found in all VHHs and have been analysed. The striking point is that 5 positions among 9 are totally conserved and only two have distinct amino acid kinds (position 1 with aromatic and positively charged residues, and position 3 with positively charged, polar uncharged and breakers). Two points must be underline here: (i) at first, these results have nothing to do with a potential problem of the PCR primer region. Indeed, it is mainly after the analysed positions [80]; (ii) the number of number of J segments is limited (e.g. 6 in llama) and so have not really provided a strong divergence between the FR4s.

The high number of structural variant patterns (19 corresponding to about two third of the total number of FR4) encompasses a larger twisted conformation directly guided by the CDR3. FR4 first positions are not flexible region, so it underlines the impact of CDR3 to guide its local conformation. Here again, no correlation between the variant patterns in terms of PBs and amino acid contents.

## Discussion

In the recent years, antibody uses have shown an impressive success for biotechnological and biomedical applications [5]. Use of bioinformatics approaches is an essential tool for engineering proteins and *a fortiori* for antibodies. A good knowledge of the sequence − structure relationship, which controls the protein folding, is essential. This is especially true for antibodies for which modifications are important questions. Humanization technology was fundamental for the remarkable progress of antibodies use for therapeutic area [81]. This optimisation needs a small set of variants. These variants design are based on the antibody structure and/or sequence information, and could impact folding, fold, fold stability and specificity. A recent *in silico* approach to perform the crafting of frameworks to accommodate other CDR regions had been proposed [82]. It combines homology modelling with simulated annealing to humanize mouse antibodies using computationally derived antibody homologous structures. However, *in silico* approaches for antibodies design for drug discovery have numerous drawbacks [83].

Compared to others antibodies, camelid VHHs have a short but rich story of experimental laboratory usage, biotechnology and biomedical purposes, *e.g.* nanobody-based cancer therapy of solid tumours [84]. Indeed, clear advantages of nanobodies compared to conventional antibodies include their size, solubility and stability. Because of these characteristics, nanobodies can be formulated as a long shelf-life, ready-to-use solution [85].

Nonetheless, in terms of structure, few works have analysed their characteristics. Many studies have been done on a limited number of VHH structures. For instance, a potential universal VHH framework was tested to graft various loops of VHHs of subfamily 2 (representing 75% of all antigen-specific VHHs), but only 5 chimeras were tested [86]. Similarly, the convexity of paratope defined by CDRs of VHHs was mainly extracted from an analysis of eight VHHs binding lysozymes [87].

Comparative molecular modelling had shown the difficulty to propose pertinent structural model of VHHs [26, 35-36], even with very sophisticated approaches as RosettaAntibody [71]. In fact, the results can be appreciated as less efficient than for classical antibodies as tested in AMA I and AMA II [88-91].

From our experiences, we have underlined a precise characteristic of VHH modelling [36], which is the difficulty to select a correct series of structural templates. At the first sight frameworks seem all highly stable, and we need mainly to focus on CDRs. In this study, we have done the first attempt to analyse precisely the structural diversity of FRs of VHHs. An important methodological asset is the use of Protein Blocks. PBs are a very useful tools as they allow a local comparison with higher precision than classical secondary tools, and are very useful to align protein structures [51, 92] and analyse protein flexibility [62, 93].

Each of the four FRs is associated to a main structural pattern, which can be considered as canonical, but represents only 40, 84, 63, and 38%, leading to the characterization to variant patterns ranging from 6 to 19. Among some of them, we observe some similarity and the final number of variant patterns could be slightly reduced. However, clearly some are really outliers at more than 1 Å from the main structural pattern and could have strong impact of the proposition of structural models. The molecular modelling performed on these extreme cases show that the structural models obtained are of low quality even for the FR regions (*data not show*). This result is in agreement with previous modelling [71]. It is also linked to the amino acid conservation of numerous FR positions, which is sometimes limited. In addition we underline that no correlation is found between the local conformational change and amino acid composition. These results indicate that long-range interactions affect the local conformation of this constrained topology and have strong implication (i) for comparative structural modelling and (ii) for antibody informatics for drug discovery. We have already observed a direct effect on the first point with some dedicated examples.

## Acknowledgments

We would like to thanks Nicolas Shinada, Akhila Melarkode Vattekatte, Jean-Philippe Meyneil and Jean-Christophe Gelly for fruitful discussions, and Pierrick Craveur for his help with the VHH structures. This work was supported by grants from the French Ministry of Research, University Paris Diderot, Sorbonne Paris Cité, French National Institute for Blood Transfusion (INTS), French Institute for Health and Medical Research (INSERM). AdB also acknowledges the Indo-French Centre for the Promotion of Advanced Research / CEFIPRA for collaborative grants (numbers 5302-2). This study was supported by grants from the Laboratory of Excellence GR-Ex, reference ANR-11-LABX-0051. The labex GR-Ex is funded by the programme "Investissements d'avenir" of the French National Research Agency, reference ANR-11-IDEX-0005-02. Calculations were performed on an SGI cluster granted by Conseil Régional Ile de France and INTS (SESAME Grant).

## Legends

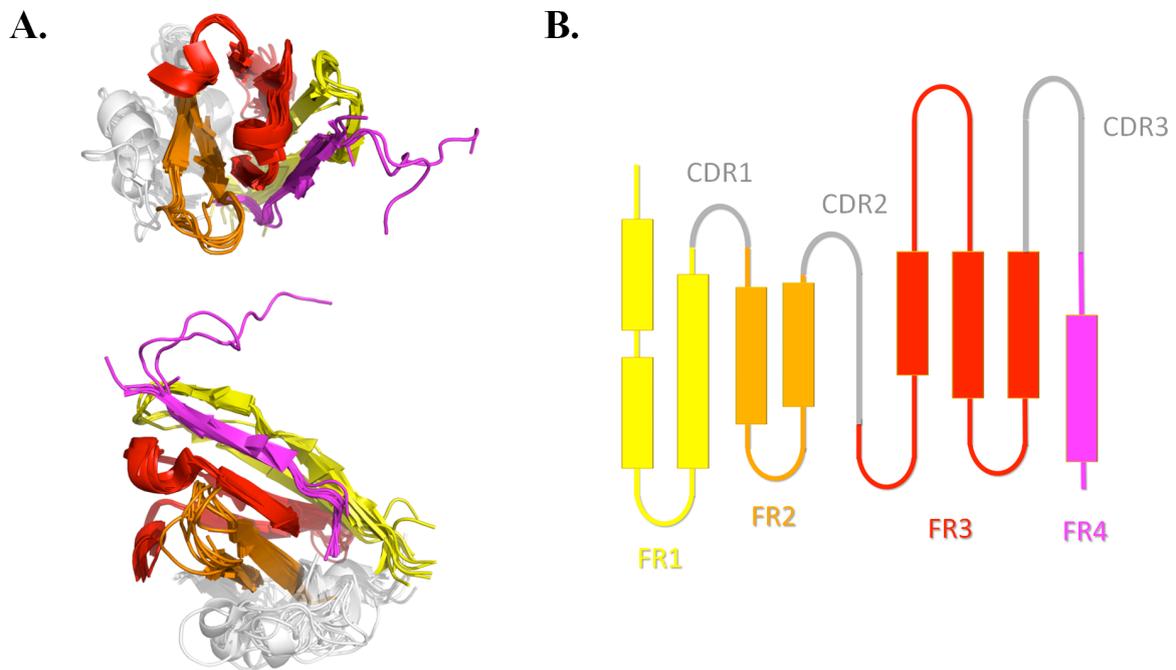

**Figure 1.** *VHH topology*. **A**) Superimposition of 10 representative VHH structures with two orientations. These 10 VHH represents the most diverse VHHs of the dataset in terms of sequences (PDB codes 3QXU chain D [8], 4GFT chain B [9], 3V0A chain C [10], 4C57 chain C [11], 4BFB chain E [12], 3CFI chain I [13], 4C58 chain B [11], 1KXQ chain F [14], 2X6M chain A [15], 2WZP chain E [16]). **B**) A 2D projection of the same information. In grey are indicated the three variable CDRs, while FR1 is in yellow, FR2 in orange, FR3 in red and FR4 in magenta. Superimposition was done using mulPBA [51] and visualisation with PyMOL software [48].

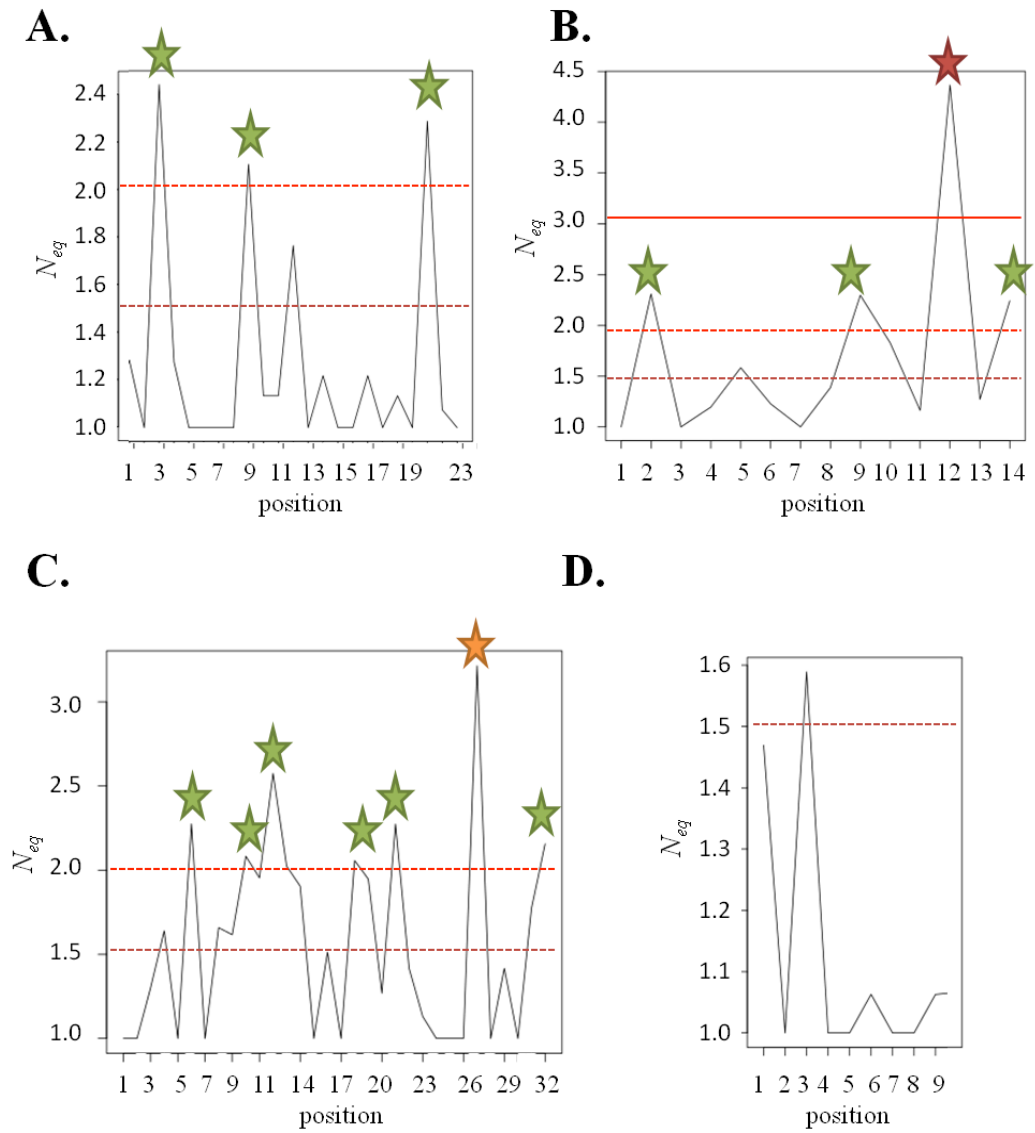

**Figure 2.** *Amino acid $N_{eq}$ distribution of the four 4 FRs*. **A.** FR1, **B**. FR2, **C**. FR3 and **D**. FR4. Stars represent $N_{eq}$ higher than 2.

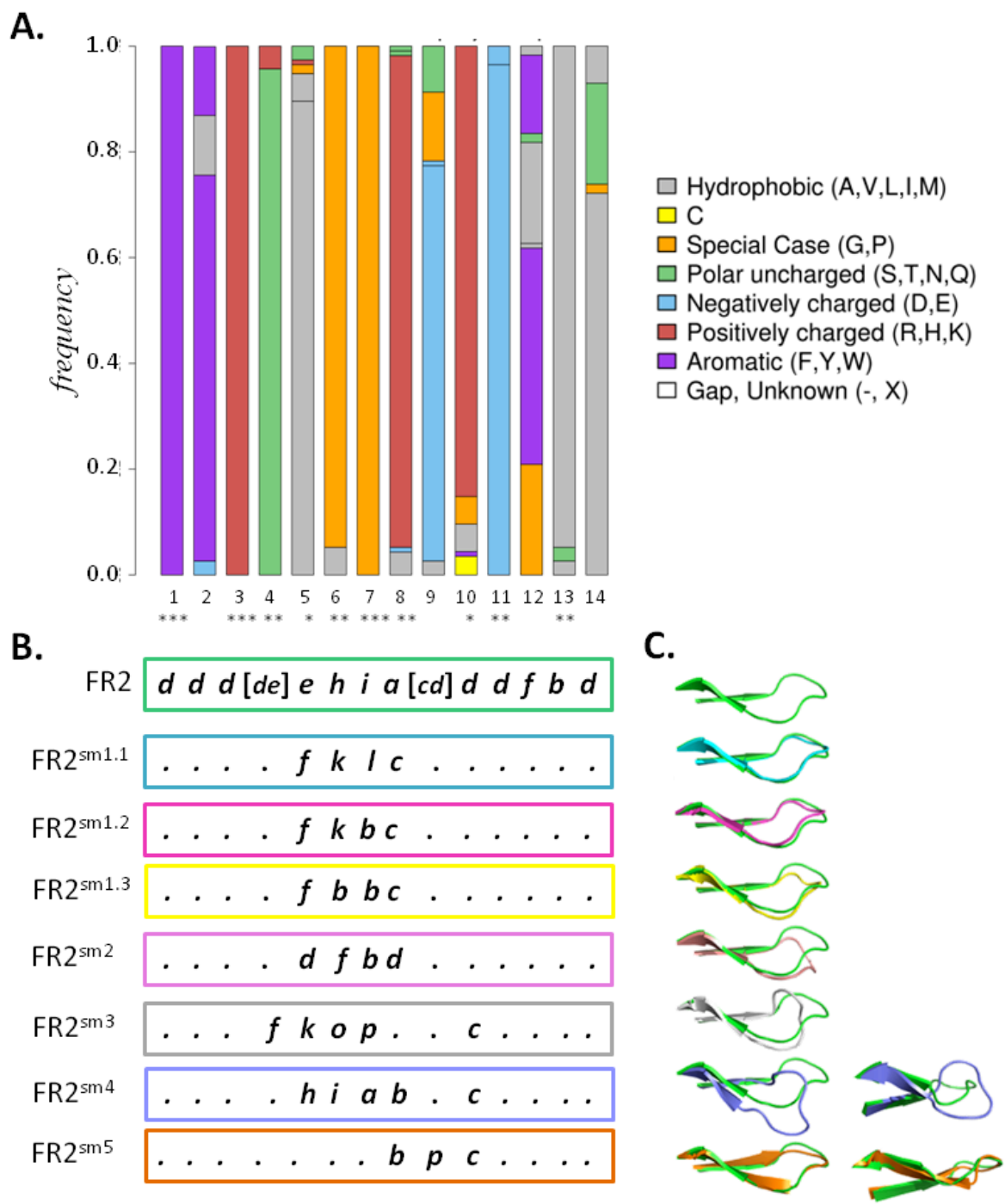

**Figure 3.** *FR2 characteristics*. **A**. amino acids frequencies at each position; **B**: Protein Blocks Representation of different structural variants of the FR2 (the dots correspond to the sequence identity with the main structural pattern); **C**. 3D representation of structural variants of the FR2. Following IMGT numbering [69-70], it corresponds to positions 36-49 which defines in IMGT this FR.

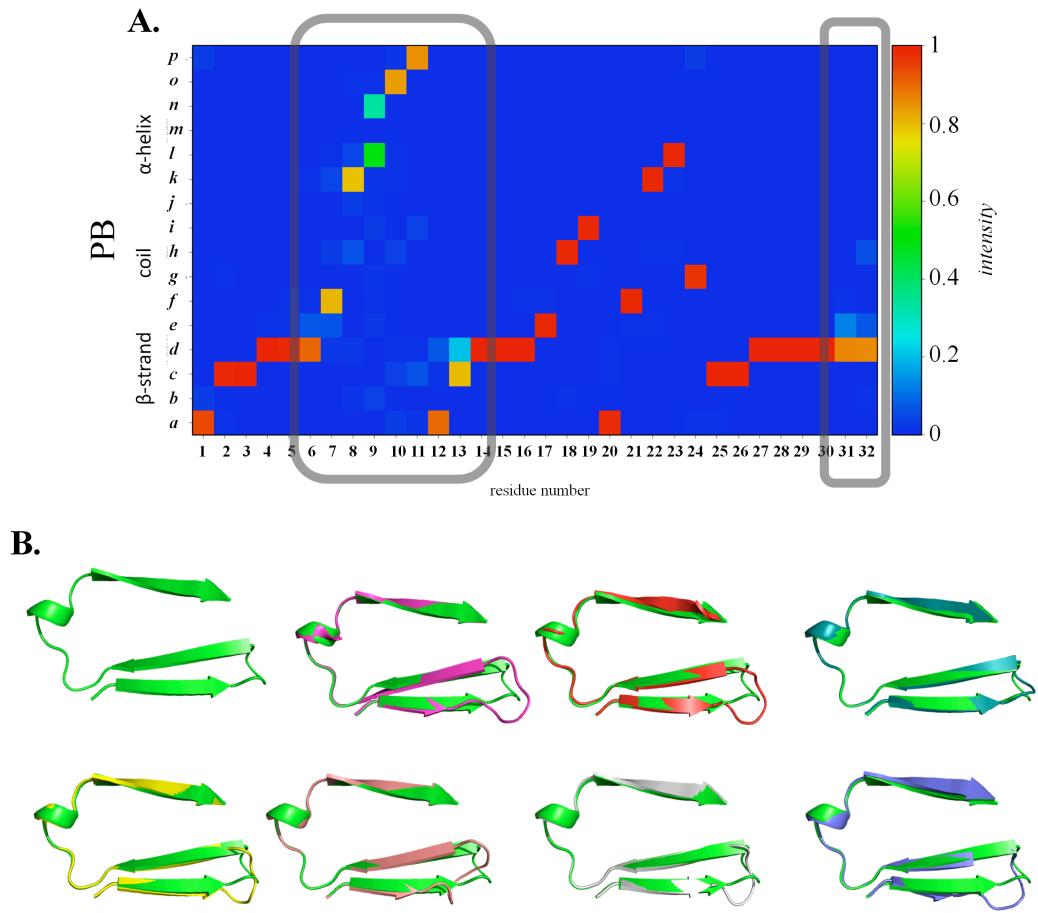

**Figure 4.** *FR3 characteristics*. **A**. PB frequencies along FR3, on the right is provided the corresponding gradient, the two more variable regions in terms of PBs are bordered in grey; **B**. 3D representation of the main structural pattern and some structural variants of the FR3.

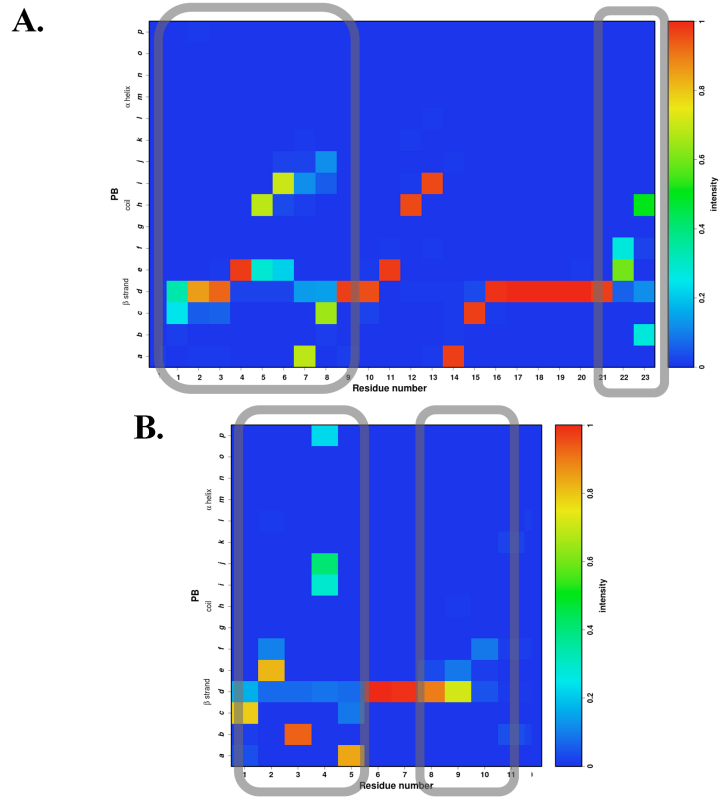

**Figure 5.** *FR1 and FR4 characteristics*. PB frequencies along **A**. FR1 and **B**. FR4, on the right is provided the corresponding gradient, the more variable regions in terms of PBs are bordered in grey.

| Region | #positions | Main pattern (%structures) | #positions with unique AA | #variant patterns (%structures) |
|--------|-----------|---------------------------|--------------------------|--------------------------------|
| FR1 | 23 | 1 (40) | 11 | 14 (60) |
| FR2 | 14 | 1 (84) | 3 | 6 (16) |
| FR3 | 32 | 1 (63) | 11 | 10 (37) |
| FR4 | 9 | 1 (38) | 5 | 19 (62) |
| Total | 78 | 1 (*56*) | 30 | 39 (*44*) |

**Table 1**. *Summary of the difference variant patterns.* The following features are noted for each FR : The FR length, the percentage of structures corresponding to the main structural pattern, the number of positions associated to only one kind of amino acid, and the number of structural variants (with corresponding percentage). The last line corresponds to the summary of all FRs, in italics is provided the mean values of structure occurrences.

**SUPPLEMENTARY DATA**

**Sup Data 1.** *Protein dataset.* The 133 protein structures are described by species and types of experiments.

| Number of structures | X-ray diffraction | Solution NMR | Total |
|---|---|---|---|
| *Camelidae* | 1 | 0 | 1 |
| *Camelus dromedarius* | 32 | 0 | 32 |
| *Lama glama* | 93 | 1 | 94 |
| *Vicugna pacos* | 6 | 0 | 6 |
| **Total** | 132 | 1 | 133 |

**Sup Data 2.** *Translation of a VHH structures in a sequence of Protein Blocks*. On the left is the 3D structure of a camelid VHH [which is in complex with porcine pancreatic α-Amylase, http://www.rcsb.org/pdb/explore.do?structureId=1kxq]. The protein is encoded as a series of (φ, ψ) dihedral angles. Each consecutive fragment of five residues is compared to the series of (φ, ψ) dihedral angles of the 16 canonical Protein Blocks [1-3]. The one with the minimal distance is associated to the central residues. The protein fragments are overlapping. Please note that the final PB sequences are 4 residues shorter than the amino acid sequences (corresponding to the *ZZ* at N and C termini).

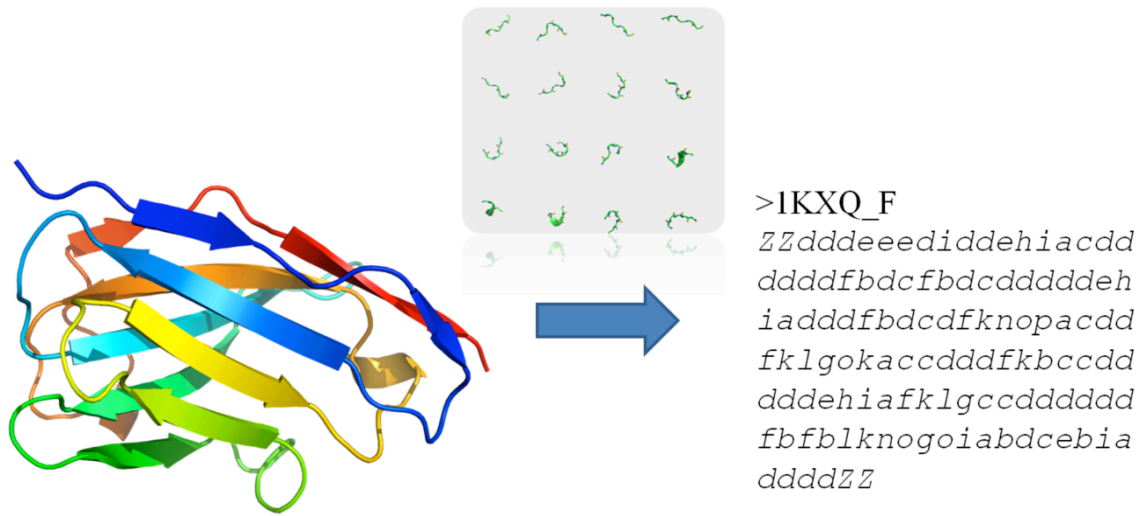

>1KXQ_F
*ZZdddeeediddehiacdd
ddddfbdcfbdcdddddeh
iadddfbdcdfknopacdd
fklgokaccdddfkbccdd
dddehiafklgccdddddd
fbfblknogoiabdcebia
ddddZZ*

**Sup Data 3.** *Redundancy of the dataset.* Identity (%ID) mean rates for each region (FRs and CDRs) between 133 VHHs sequences obtained from the PDB (and associated standard-deviation, sd).

|  | % ID (mean) | % ID (sd) |
|---|---|---|
| FR1 | 88.70 | 4.06 |
| FR2 | 76.91 | 8.54 |
| FR3 | 79.35 | 4.51 |
| FR4 | 94.43 | 4.70 |
| CDR1 | 33.47 | 10.32 |
| CDR2 | 52.13 | 6.19 |
| CDR3 | 19.31 | 4.32 |

**Sup Data 4.** *FR3 characteristics*. **A**. Amino acid $N_{eq}$ of FR3. **B**. Occurrence of amino acids at each position. Following IMGT numbering [4, 5], it corresponds to positions 66-94.

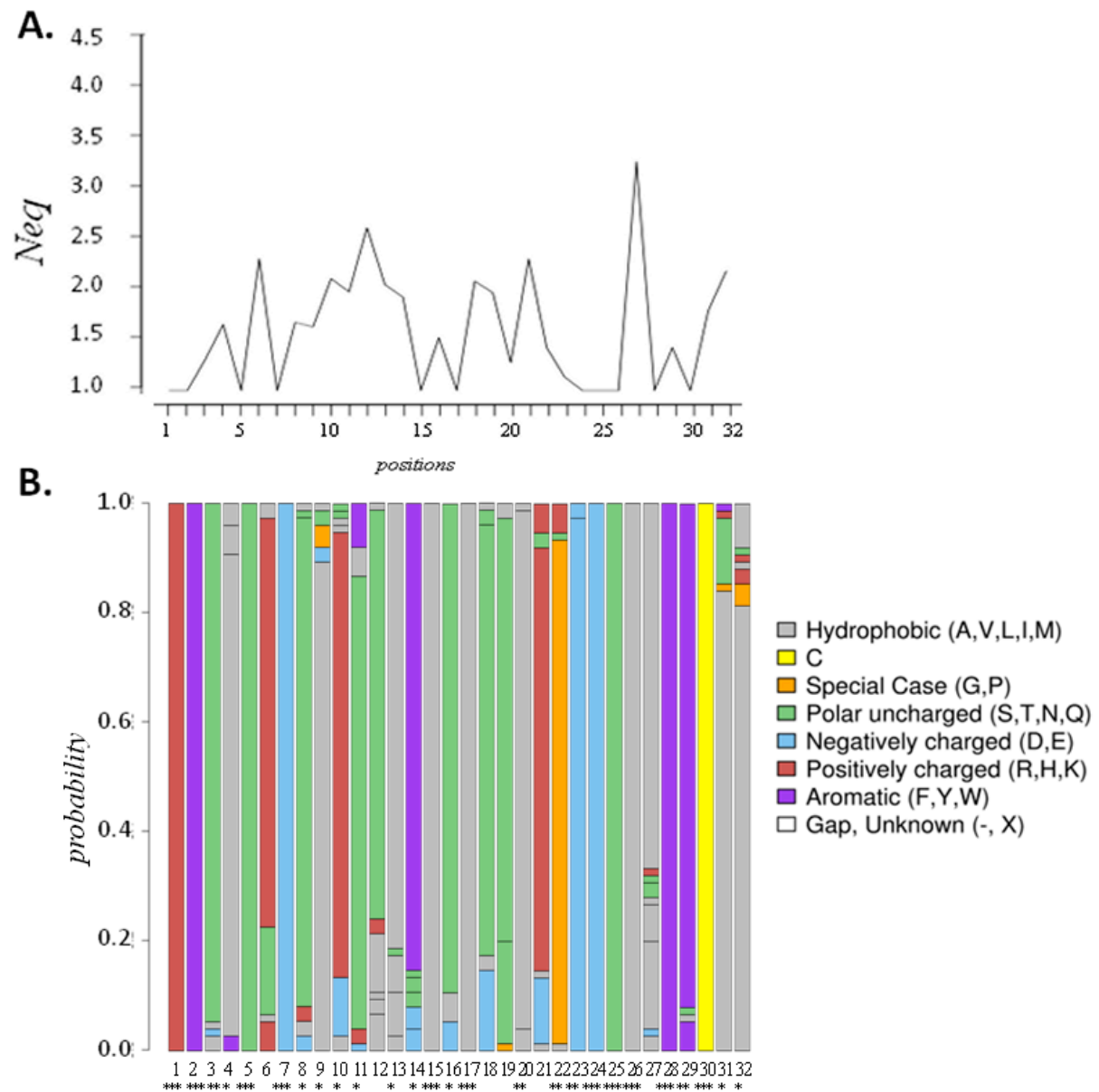